\documentclass{ws-procs975x65}
\usepackage{graphicx}
\usepackage{amsmath}

\def\beq{\begin{equation}}
\def\eeq{\end{equation}}

\begin{document}

\title{Study of Accretion processes Around Black Holes becomes Science: Tell Tale Observational Signatures of Two Component Advective Flows}

\author{Sandip K. Chakrabarti$^*$} 

\address{S.N. Bose National Centre for Basic Sciences, JD-Block\\
Kolkata, West Bengal 700106, India\\
$^*$E-mail: chakraba@bose.res.in}
\address{Indian Centre for Space Physics, 43-Chalantika, Garia Stn. Rd.\\
Kolkata, West Bengal 700098, India\\ 
}

\begin{abstract}

An accretion flow around a black hole has a saddle type sonic point just outside the event horizon to guarantee that the flow enters
the black hole supersonically. This feature exclusively present in strong gravity limit makes its marks in every observation of black hole candidates. 
Another physical sonic point is present (as in a Bondi flow) even in weak gravity. Every aspect of spectral or temporal properties 
of every black hole can be understood using this transonic or advective flow having more than one saddle type points. 
This most well known and generalized solution with viscosity and radiative transfer has been verified by numerical simulations also. 
Spectra, computed for various combinations of the standard Keplerian, and advective sub-Keplerian components 
match accurately with those from satellite observations. Standing, oscillating and propagatory 
oscillating shocks are produced due to centrifugal barrier of the advective component. 
The post-shock region acts as the Compton cloud producing the power-law spectra. Jets and outflows are also 
produced from this post-shock region, commonly known as the CENtrifugal barrier supported BOundary Layer or CENBOL. 
In soft states, the CENBOL is cooled down by soft photons from the Keplerian disk, and thus the outflow 
is absent. Type-C and Type-B QPOs are generated respectively due to strong and weak resonance oscillations 
of the CENBOL. Away from resonance, oscillation may be triggered when Rankine-Hugoniot conditions are not satisfied and Type-A QPOs could be seen.

\end{abstract}

To be published in Proceedings of 14th Marcel Grossman meeting, Eds. R. Ruffini, R. Jantzen, M. Bianchi, (World Scientific: Singapore), In press.


\bodymatter

\section{Introduction}

In the past, the subject of accretion flows has been seen to be as good as meteorology. Unwarranted assumptions and simplifications 
marred the subject in a way that it is assumed that observations have to be understood by phenomenology alone -- as though
no physical theory could possibly be good enough to enable understanding timing and spectral properties 
under a common framework. The situation was similar to what we had in early days of gamma-ray burst models, or even the models of cosmology,
people could construct any odd model and draw any odd picture and pass it as the best model with a new name.
Even a self-similar flow solution which is fundamentally wrong in black hole astrophysics would be considered to be refreshing and a worthy contribution.
Fortunately, in the last decade, the situation has changed dramatically. Today, it gives a great confidence to the community
when detailed spectral and timing features can be explained using results based on a specific 
theoretical solution. The subject has finally become a part of `Science'. A Bondi flow was rejected outright 
due to its inefficiency of emission. When there was no other solution available, the prediction of a simple 
multi-color black body radiation bump from a standard Shakura-Sunyaev (hereafter SS73) disk\cite{ss73} disk was welcomed 
by the galactic or extragalactic black hole community. However, today the observers are amazed by the complexity 
of the spectral and temporal variation and these could no longer be explained by the standard disk. As a result, many 
phenomenological models were advanced, keeping the standard disk as the basic solution. 
This generated too many patch up models. People were not willing to give up the standard disk. 
With the advent of generalized transonic or advective solutions in the 1980s and 1990s\cite{c89a,ac90,c96a} 
and the success of its explaining spectral transitions by variation of accretion rates 
(\refcite{ct95}, hereafter CT95) it became clear that the standard disks, with an inner edge at the marginal stable orbit, 
are no longer necessary to explain any feature about black holes. The generalized
two component advective flow solution of Chakrabarti\cite{c89a,c89b,c90,c96a} and its spectra \cite{ct95,c97} 
contain a version of the standard disk as a special case, 
which does not stop at the marginally stable orbit but directly reaches the black hole horizon supersonically. 
Similarly, advection dominated flows, such as the Bondi flow, is also a special case of this solution. 
The so-called Advection Dominated Accretion Flow (ADAF)\cite{ny94}
is a self-similar solution of the same generalized flow, though in the context of black hole accretion it is not useful, since
it has no length scale, and thus, for instance, it does not have any sonic point to 
enter into the horizon, or the shock transition (natural Compton cloud), so vital to explain the observations of any black hole candidate.

Since transonic flow solution is of very rigid nature, number of free parameters are fixed and the model is not flexible. 
This is the best part which elevates the subject to the pedestal of `Science'. TCAF requires only five physical parameters: i) The mass of the black hole,
ii-iii) the two accretion rates (one of viscous optically thick component ${\dot m}_d$ and the other of sub-Keplerian optically thin,
geometrically thick, halo component ${\dot m}_h$); iv) the 
location of the centrifugal pressure supported shock and the compression ratio (ratio of densities 
of the post- and pre-shock region). This shock is produced when the Rankine-Hugoniot condition is satisfied and the post-shock region 
is the CENtrifugal pressure dominated BOundary Layer or CENBOL which is the so-called hot-electron cloud, or Compton cloud which passes 
on its thermal energy to the soft photons coming from the Keplerian disk.
In TCAF, the CENBOL boundary (shock) is the outer boundary of the Compton cloud 
and the inner boundary of the Keplerian disk. The accretion rates
decide density distribution and the compression ratio decides the optical depth of the Compton cloud. So each parameter has multiple functions, which is 
why number of parameters is minimum. Of course, there could be models such as power-law + disk black body which may be able to fit the 
spectrum equally well, but that good fit does not give any physical quantity about the system, other than the measured flux. On the contrary, TCAF
fits give us the physical parameters and in outburst sources, the evolution of the parameters.   

\section{TCAF Paradigm}

\subsection{Spectral state transitions}

Chakrabarti\cite{c90, c96a} pointed out long ago that if the viscosity parameter is higher than certain critical value then only
a flow injected very far away would become a Keplerian disk (also see, \refcite{nc16}). 
In reality, this is a statement on the viscous stress involved in transporting angular momentum.
A Keplerian disk with certain accretion rate requires certain minimal stress to maintain itself.
The Keplerian disk formation was seen through numerical simulations in one dimension\cite{cm95} and
in two dimensions\cite{gc12,gc13,gc15}. 
In this case, centrifugal pressure supported shocks are produced first in the post-shock region where the 
shear stress is the highest (as the pressure is high) and then propagate outward while converting 
the sub-Keplerian flow into a Keplerian flow. Thus, if a low angular flow is injected into the system 
which becomes Keplerian only along the equatorial plane due to higher viscosity, the flow will 
be de-segregated into two components (CT95) and depending on the relative importance of the Keplerian 
rate with respect to the sub-Keplerian halo, the CENBOL may or may not be cooled down. 
When the Keplerian rate is high, it cools down the CENBOL and a soft-state is produced. If the Keplerian rate
is low, it does not cool the CENBOL and we have spectrally hard states (CT95).  

This findings are the turning points of the entire subject of accretion flows. This understanding agrees totally what is grossly observed: 
objects are generally in hard states, and they go to soft-states for a short duration. In outburst sources, soft-states are achieved only 
when viscous Keplerian matter rushes towards the black hole. In some outbursts the soft states are formed since 
the stress is never high enough. Thus, according to the TCAF paradigm, the standard SS73 type Keplerian disks are 
exceptions rather then the rules in systems involving black holes. 

\begin{figure}[h]
\begin{center}
\includegraphics[width=.65\textwidth]{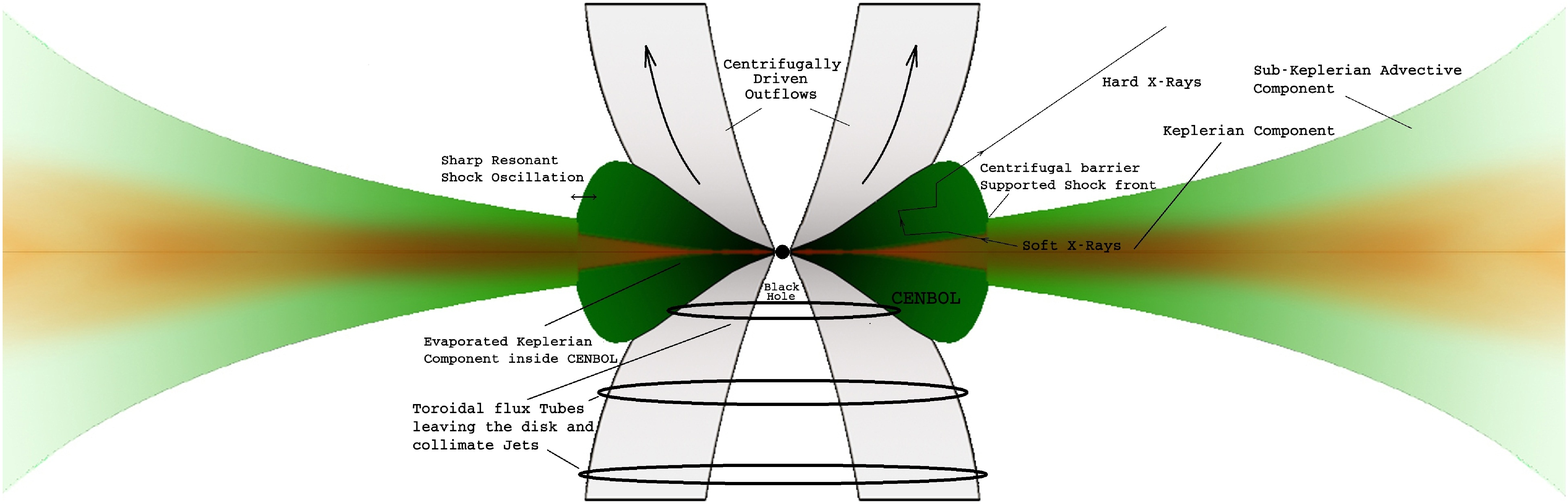} \\
\vspace{1.0cm}
\includegraphics[width=.65\textwidth]{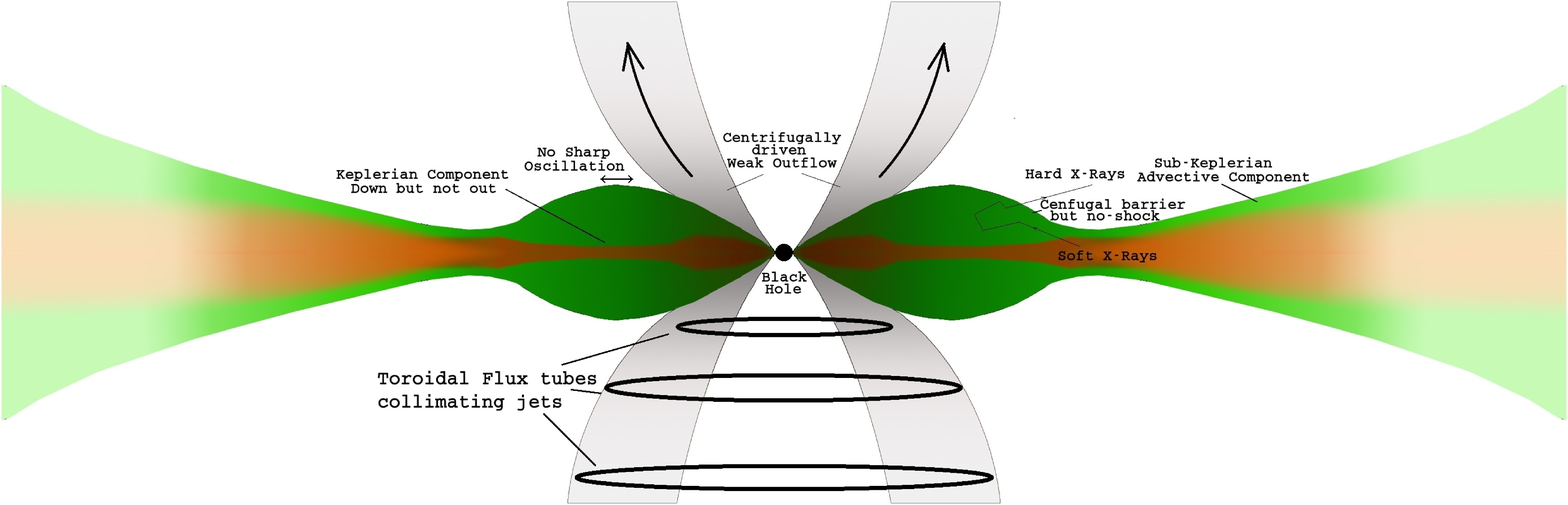} 
\caption{A schematic picture of the accretion and radiation processes in a Two Component Advective Flow (TCAF) solution with a centrifugal barrier supported
shock front close to a black hole (top) and a schematic solution when the shock condition is not satisfied (bottom).}
\end{center}
\end{figure}  

Figure 1a describes a TCAF flow which includes a shock where the flow is tenuous and supersonic only before the shock and in a narrow region
outside the horizon. It is the most general transonic flow. Figure 1b does not include a shock as the shock condition is not
satisfied. However, it has the centrifugal barrier where the flow remains tenuous and supersonic{\cite{cn00,c97}}. 
The presence of the CENBOL, or the shock is possible since the black hole allows the flow to have an 
innermost sonic point so that the flow can become supersonic even after a shock transition 
(super-sonic to subsonic). In a Newtonian flow, such a shock could form only on the surface of a star and the post-region becomes the 
normal boundary layer (BL). In analogy we have also been calling the post-shock region to be the BL of a black hole (CENBOL).
Difference is that in a neutron star, the thickness of BL is a few meters while in a black hole, the CENBOL could be a few tens to hundreds of 
Schwarzschild radii. The previous incarnation of CENBOL was a thick accretion disk \cite{ajs78, pw80}
which is really a non-advecting optically thick, radiation pressure dominated generally sub-Keplerian, 
disk. Another incarnation was the ion pressure supported torus \cite{r82}
which is hot and optically thin torus. CENBOL is naturally produced in the post-shock region. It puffs up because the post-shock 
flow is hotter than the pre-shock flow. It becomes geometrically thick for both very high and moderate to low accretion rates in the 
Keplerian components. When the Keplerian disk rate is high enough and it cools the CENBOL, one can expect that a Keplerian disk 
would form. However, the difference of this disk from its earlier incarnation, namely, SS73 disk is that our solution is advective,
it passes through the inner sonic point at around $\sim 2.5$ Schwarzschild radii, where the inner sonic point is located (for a non-rotating black hole). To pass through this point the flow
prepares to become sonic from $r>r_{ms}$, where $r_{ms}$ is the marginally stable orbit.
In contrast, SS73 has its inner edge permanently fixed at $r_{ms}$, where the flow is still Keplerian and sub-sonic. Figure 2 is a general TCAF configuration
around a neutron star. The magnetic field is moderately strong and halts most of the flow and diffuse it along the 
field lines creating a magnetosphere. Here, matter can directly hit the star surface and create a Boundary Layer (BL). Comptonization of soft photons
from the disk, from the BL as well as synchrotron radiation from the magnetic field is thus possible by the CENBOL and the magnetosphere. The CENBOL
does not have to be present in the neutron star case. The spectrum is thus far more complex, in general, though in special cases it may look 
like that of a black hole.

The shock solutions which were explored by Chakrabarti \& collaborators have been independently verified by several workers. Recent works \cite{l16}
show that standing shocks are formed in ADAF also when the equations are solved using procedures as given in Chakrabarti\cite{c89a,c89b}, 
though isentropic shocks may be unstable in some region of the parameter space.

\begin{figure}[h]%
\begin{center}
\includegraphics[width=.65\textwidth]{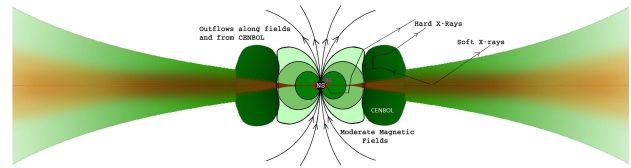} 
\caption{A schematic diagram of accretion dynamics and radiation processes in a Two Component Advective Flow solution with a centrifugal barrier supported
shock front close to a neutron star with moderate magnetic field. The Boundary Layer (BL) on the star surface as a result of  a shock,  
and a secondary Compton cloud in the magnetosphere apart from the possible presence of CENBOL makes the spectra complex.}
\end{center}
\end{figure}  

Generalized transonic flow was already considered to be the most general 
type accretion flows even in 1993. The following quotation from the UNAM 
conference (1993) proceedings is in order: {\it There are increasing observational evidences that accretion disks in nature are not just thin, thick, slim or fat, smooth or shocked, twisted or centrifugally exhausted over their entire lifetime .... What shape and form a disk might assume depends on the energy, angular momentum and the quantity of matter 
supplied, as well as dominant viscous mechanism that may be prevalent.....  This view has a unifying character, but details are still to 
be worked out using both the approaches we discussed in this review}\cite{c94a}.
 Since then, the road map given in the review has been followed
and as a result the subject is in a satisfactory shape.  This grand unification of all possible disk solutions 
and the unified view has been emphasized again and again with more compelling evidences 
\cite{c94b,c96b,c98a,c98b,c98c,c00,c01,c02a,c02b,c07,c08,c09,c10,c12,c13,c15,c15minsk}.

\subsection{Quasi-periodic oscillations}

Discussion above focused on the spectrum only. However, Molteni, Sponholz and Chakrabarti (\refcite{msc96}, hereafter MSC96) discussed that when the infall 
time scale in the CENBOL roughly agrees with the cooling time scale, a resonance oscillation sets in. A power-law cooling was considered 
for simplicity. This was later extended for a stronger cooling akin to Comptonization\cite{cam04} and eventually with actual Comptonization\cite{g14}. 
In all the cases, the resonance oscillation is found to cause the low frequency QPOs. In Chakrabarti et al.\cite{cetal15}
theoretical justification of the resonance was given, it was further shown that in outburst sources, 
once the resonance sets in, it would continue to be locked into resonance till accretion rates change considerably, causing evolution of QPOs 
in the rising and declining phases. Most interestingly, this condition allows one to obtain QPO frequency from the shock location or 
the size of the Compton cloud (CENBOL). Thus a firm connection between the spectral and temporal properties was established. 
QPO evolution has been explained to be due to systematic evolution of the Compton cloud size in the so-called 
propagating Oscillatory shock (POS) of Chakrabarti and collaborators\cite{cetal05,cetal08,detal10,detal13,dc10,netal12}.
Molteni et al.\cite{metal99} and Okuda et al.\cite{oetal07}
showed that the axisymmetric shocks remain stable even after 
strong non-linear non-axisymmetric perturbations. These non-axisymmetric blobs attached to the centrifugal barrier supported shocks 
can also induce QPOs of exactly the same frequencies as does the vertical oscillations of Chakrabarti et al. \cite{cam04}).
Nagakura \& Yamada\cite{ny08} concludes that these non-axysymmetric blobs render the axisymmetric shocks unstable which is incorrect.   

\subsection{Jets and outflows}

In TCAF paradigm, jets and outflows are strictly coupled to the CENBOL. CENBOL is the base of the outflow as long as it remains hot\cite{mlc94,c98a,c98b,c99,getal12}.
The outflow is not collimated but the matter supply is sufficient. It's acceleration must be done by separate mechanism (if required) either by radiative processes\cite{chatetal02,chatetal04}
 or by sudden collapse of magnetic field lines in CENBOL\cite{netal01}.
 There is no need of any magnetic fields 
in explanation of spectra or timings of the radiation from the disks. 
So the magnetic field does not seem to be dynamically important in accretion disk physics. 
A corollary is that mechanisms such as in Blandford-Znajek \cite{bz77} which solely bank on magnetic energy extraction are irrelevant. In any case, those mechanisms require large scale aligned field lines coupled to the disk and extraction of energy is coupled 
to the spin of the black holes. Furthermore, such mechanisms do not distinguish between hard and soft states, 
while in reality, outflows are prominent only is hard states. This is achieved in TCAF most naturally since CENBOL is present only 
in harder states. Once CENBOL is cooled down in soft states, normal production mechanism of jets is stopped automatically. This is precisely what is observed. 
Chakrabarti\cite{c89a,c89b,c99} presented the first Global Inflow and Outflow Solution (GIOS) where a formalism 
is presented to compute the outflow rate from the inflow rate theoretically.
Das \& Chakrabarti \cite{dc99} and Das et al. \cite{detal01} extended this for flows with rotation.
It was found that the shock strength is intimately related to the outflow rate. When the shock is non-existent 
(Compression ratio $R= 1$), the outflow rate is zero. For strongest shocks also the rate is lower. 
Only for intermediate shock-strengths ($R\sim 2-3$) outflow rates are the highest. 
In enigmatic objects such as GRS1915+105 the shock strength decides the variability state 
resulting in variation of Compton cloud geometry\cite{pc13,pc14,pc15}. Theoretical works mentioned 
above have been verified by detailed Monte Carlo simulations with Comptonization\cite{getal12}.

In TCAF, the collimation of the jets are done by hoop stress of the toroidal field lines (Figs. 1ab). 
Because of fast rotation, even a small stray field cannibalized from the companion would be converted 
predominantly to a toroidal field. Detailed simulations\cite{cs94,sc94,dgc16}
show that thin field lines could be `smuggled' into the inner regions of the disk since the drag force is negligible. However, near CENBOL it 
escapes to the funnel area and move along with the jets, collimating them. Field lines with large cross-sections 
escape due to buoyancy far away from the black holes and eventually collimate the outflows. Sudden ejection of matter 
is possible when the tension in the field lines dominate inside CENBOL. They rapidly collapse destroying the 
CENBOL or the Compton cloud altogether. This condition can produce softer states at lower intensity 
as is seen in certain variability classes of GRS 1915+105\cite{netal01}.

\section{TCAF fits of Outbursting sources}

Encouraged by the fact TCAF has all the features which are seen through observations (and which numerous 
`models' struggle to explain on a case by case basis), we went ahead and added TCAF as an additive table model 
of the spectral fitting software XSPEC of HEARC. We wanted to target the outburst sources first as they are 
known to change the spectral states in rapid succession and we wanted to know how the physical parameters 
of the disk must have varied in order to achieve such spectral classes. Furthermore, since low-frequency QPOs 
are found to evolve on a daily basis, we can verify self consistency between the QPO frequency 
and the shock location by checking if they satisfy resonance condition.  

A large number of recent papers \cite{netal12,detal13,detal14,detal15a,detal15b,mondal14a,mondal14b,mondal15,jana16,molla16a,molla16b} demonstrate that 
TCAF is very successful in fitting these data, thereby obtaining very satisfactory evolution of the mass accretion rates in the two 
components, size of the Compton cloud and the shock compression ratio. The QPO frequency obeys the resonance condition. 
In the rising phase, the QPO frequencies rise with day as the shock moves in with the rise of the Keplerian rates. In soft-intermediate 
states, the system deviates from a resonance condition for varieties of reasons: since the accretion rate of the Keplerian component rises, the cooling time scale is reduced drastically, but the CENBOL may start expanding due to enhancement of the radiation pressure and the infall time scale rises. In this situation, the shock condition
may not be satisfied, but oscillation may follow because of the presence of the innermost sonic point\cite{rcm97}. This gives rise to sporadic Type-A QPOs.

What transpired from these fits is a kind of universality in the pattern of variations of the rates, 
and the entire episodes, objects after objects, can be understood with TCAF solution. It became clear that as 
in dwarf novae outbursts, increase in viscosity is responsible for the outbursts of transient black hole 
candidates. The sudden rise in viscosity causes a sudden increase in the Keplerian rate and sub-Keplerian rate. 
Sub-Keplerian matter, being of lower angular momentum, rushes in towards the black hole earlier than the viscous, 
Keplerian disk matter. Thus these two rates may have peaks on two different dates.
The Keplerian disk arrival time gives a direct measurement of the viscosity in the Keplerian flow. 
QPO frequency gradually rises as the soft-photons injected from the Keplerian component cool down the outer edge 
of the CENBOL and the CENBOL shrinks and oscillates at shorter time scale. In all the cases, QPO achieves a 
local maximum while transiting from the hard-intermediate to soft-intermediate states in the rising phase. 
Opposite is true for the declining phase of the outburst. A plot of Accretion rate ratio vs. 
QPO frequency gives a complete physical process very clearly, unlike the so-called `q' diagram 
of Hardness-Intensity ratio where physical basis is missing. The latter diagram looks totally different 
for different outbursts and also same hardness ratio becomes meaningless for a black hole of different masses. Even if the conclusions  are
TCAF dependent, since TCAF is a solution and not a cartoon, the plot is expected to be universal.
\begin{figure}[h]%
\begin{center}
\includegraphics[width=.85\textwidth]{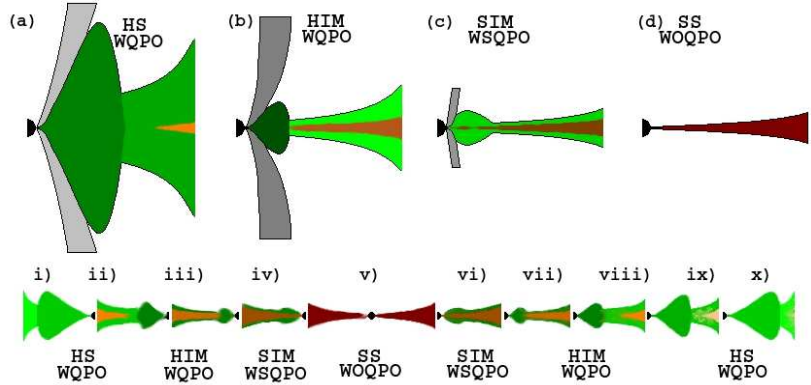} 
\caption{Schematic flow configurations in Hard State (HS), Hard Intermediate State (SIM), Soft Intermediate State (SIM) 
and the Soft State (SS) in an outbursting black hole
candidate (Top). Also shown are the typical sequence of configurations through which an entire phase of the outburst passes. The declining 
phase is more complex than the rising phase since the Keplerian disk takes a different route to disappear than the one it took 
during its formation stage (bottom). Here, W: with; WO: without; WS: With sporadic.}
\end{center}
\end{figure}  

Figure 3 shows sequence of most likely TCAF configurations of the Hard, Hard Intermediate, Soft Intermediate and Soft States as is inferred 
from the accretion rates and shock locations. In the lower panel, we show the sequence in which the rising 
phase and declining phase occur. One of the observations which is evident even from the light curve is that, 
typically, the rising phase takes a shorter time 
than the declining phase, and it is not necessary that the soft-intermediate or soft states are achieved at all. 
The explanation of the first point which manifests itself as a hysteresis effect (rising and declining phases are time asymmetric) 
is that the creation of a Keplerian disk is much faster than its disappearance. When viscosity is reduced, triggering the beginning of a declining phase, 
the Keplerian disk already formed has difficulty to drain its matter. So it hangs on in soft/soft 
intermediate states while mixing with the fast moving advective component till it withers away. However, once it is fully advected,
the time taken in hard-intermediate and hard states are shorter in the declining phase because the accretion rate of the remaining matter is drastically 
reduced. This process has been reproduced by numerical simulations and would be presented else where\cite{rc16}. Presence of two components has been verified repeatedly in the literature\cite{sm01,sm02,cs03,t14}.

While fitting with TCAF solution, we not only need the least number of parameters (five, including the mass), we do not want that the so-called normalization constant
which is required to scale the observed data with the calculated spectra to vary from day to day. We allow it to remain a constant, 
because strictly speaking, this should depend only on the mass, distance and the inclination angle only. 
However, since the mass is usually not known and we leave it a free parameter, the normalization constant also varies within a narrow 
range even when we fit data using all the states in an outburst. Our mass measurements from various outbursts of the same 
object use the same normalization constant\cite{molla16b}. However, since we measure
mass of a black hole mainly using a thermodynamic quantity such as the temperature of the gas, and since the temperature is not very sensitive to the 
mass, there is always a significant error bar in such a procedure. In other models, such as diskbb+PL, the disk inner edge is computed from the normalization, which often comes out
to be an absurd number (inside the black hole). This inconsistency is due to the simplistic computation of the disk contribution and its Comptonized component.

Another important fact about fitting with TCAF is that one does not require an additional reflection component as in diskbb+powerlaw model, for instance. This is 
because we do not assume a power-law to begin with and thus we do not need to correct this assumption by adding a reflection. Our spectrum already includes the 
reflection component produced by the radiation from CENBOL on the cooler Keplerian disk. Presently, TCAF solution is not equipped with fitting the line emissions 
starting from abundances of the metals in the companion, and one requires to add a Gaussian or `laor' model along with TCAF to fit unresolved lines or double lines
as the case may be. This work is in preparation and would presented elsewhere. 

\begin{figure}[h]
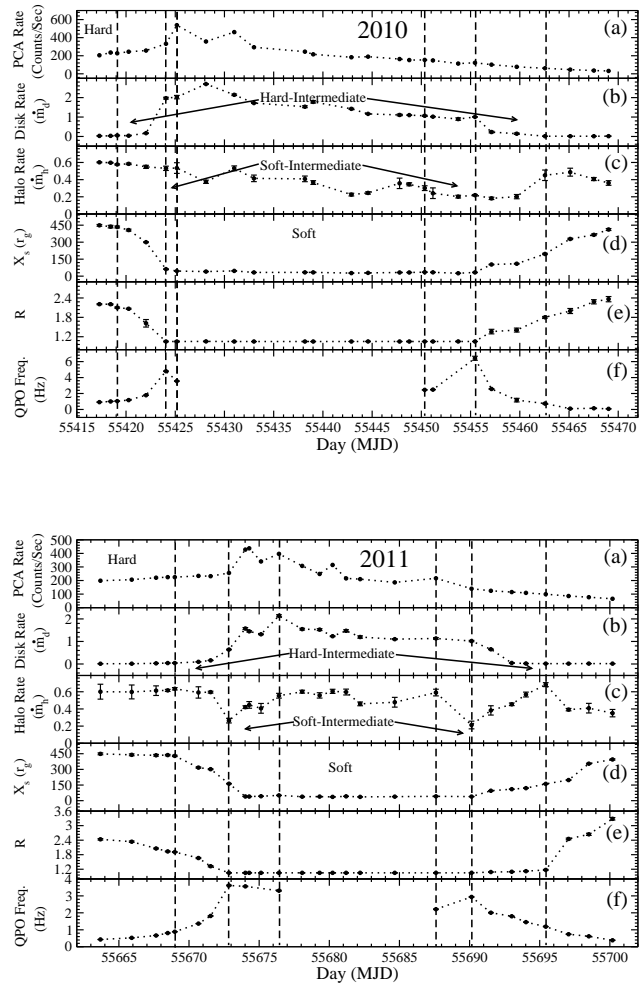
%
\begin{center}
\includegraphics[width=.65\textwidth]{fig4a.eps} \\
\vspace{1.0cm}
\includegraphics[width=.65\textwidth]{fig4b.eps} 
\caption{Evolution of TCAF fitted parameters in (a) 2010 and (b) 2011 outbursts of the black hole candidate H1743-322. Other than the 
observed quantities such as the PCA count rate and QPO frequencies, plots of disk accretion rate (${\dot m}_d$), halo accretion rate (${\dot m}_h$),
shock location ($X_s$) and the compression ratios are also shown (From Molla et al. 2016).}
\end{center}
\end{figure}  

Figures 4(a-b) give typical variations of physical quantities during two complete outbursts for H1743-322 taken from Molla et al.\cite{molla16b}.
This is generally the trend. After the outburst is triggered, the halo rate either remains almost 
constant and sometimes may go up. After a few days the Keplerian rate goes up. 
The shock location (or, the size of the Compton cloud) is reduced monotonically and the shock strength is also reduced as the Keplerian rate rises. 
In the soft-intermediate and soft states, the shock is virtually absent, only the rates vary. As the viscosity at the outer edge is reduced, 
triggering the beginning of the decline phase, the states are visited in reverse order. The shock location rises again with its increasing 
strength and the disk component rates also start to decrease. Since the disappearance of the Keplerian component in the absence of viscosity 
is a slower process, through mixing with the advective component, for instance, the decline phase is not time symmetric. 
Results of numerical simulations will be presented elsewhere\cite{rc16}.

So far, we used only the low energy data from RXTE or nuStar satellites while fitting. High energy data
could be fitted with TCAF equally well, provided we incorporate Comptonized photons from the 
non-thermal electrons generated through shock acceleration at the boundary of the CENBOL. Such 
non-thermal electrons were injected by hand before\cite{wz01},
but our solution produces these electrons from CENBOL self-consistently\cite{cm06,mc13}.

\section {Time lag properties}

TCAF is not only successful in explaining the spectral properties and QPOs, the solution can explain why time lag changes with QPO frequencies and why high 
inclination angle sources which customarily exhibits harder spectra\cite{ghoshetal11,hetal15} also show switching of the time lag from positive to negative at some 
QPO frequency. Figures 5(a-b) taken from Dutta \& Chakrabarti\cite{dc16} 
show that in the low or high inclination source such as  GX339-4 (2007 outburst),  or XTE J1550-564, 
the time-lag increases with the shock location as it should be for Comptonization process, but in the latter case it switches sign due to focussing effects. 

\begin{figure}[h]
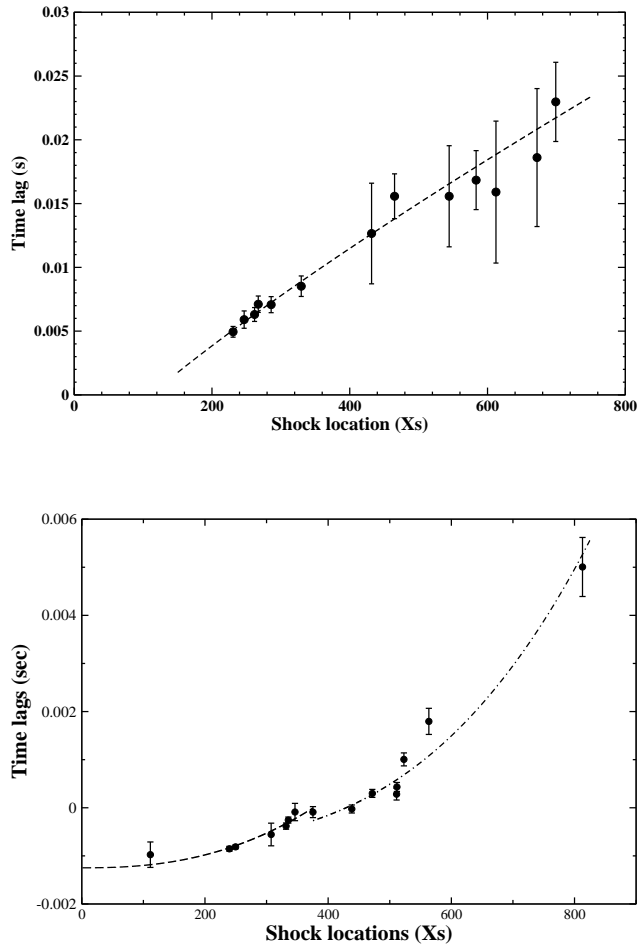
%
\begin{center}
\includegraphics[width=.65\textwidth]{fig5a.eps} \\
\vspace {1.0cm}
\includegraphics[width=.65\textwidth]{fig5b.eps} 
\caption{The computed time lag increases with the shock locations of (a) GX339-4 and (b) XTE J1550-564 as derived from the 
propagatory oscillating shock (POS) solution\cite{dc16}.
 }
\end{center}
\end{figure}  

\section{Concluding Remarks}

TCAF is a complete solution which explains spectral and timing properties of galactic and extra-galactic black hole candidates in a single `package deal'
without changing the solution itself. TCAF fits file has been able to explain all the known results and minimum number of parameters are used. This is thus
a `all or nothing' solution. There are several cartoon models in the literature which capture 
some phenomena for some time, but not all phenomena for all the time. These may also fit the data but they 
usually require larger number of free parameters due to self-inconsistencies in choice of parameters of the set. 

\begin{figure}[h]%
\begin{center}
\includegraphics[width=.65\textwidth]{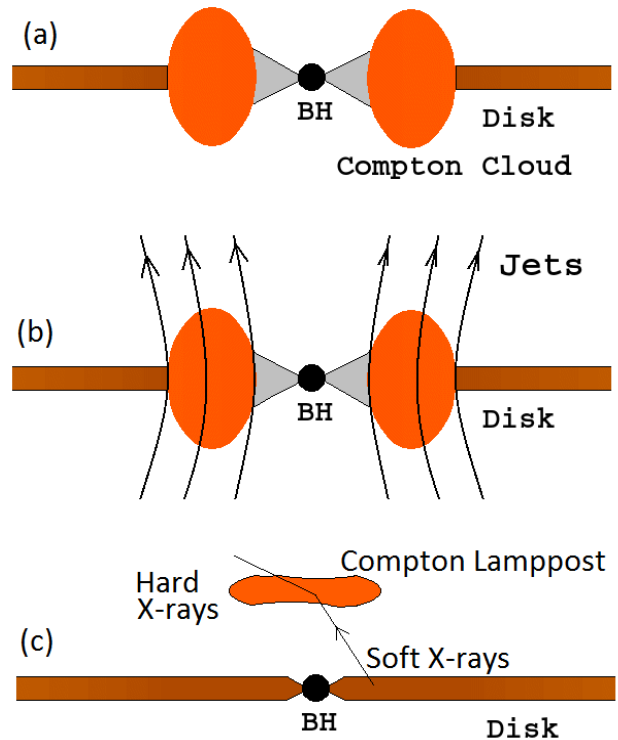} 
\caption{Examples of models which are incompatible with governing equations. (a) Disk puffs up to produce its Compton cloud - A self consistent spectral state is always soft,
collapsing the Compton cloud. (b) Magnetic fields anchored from external agency to arrest the disk flow letting strong jet formation - such a large scale aligned field configuration is 
impossible in a dynamic system. (c) Compton cloud shining as a lamp post on the disk - would introduce too many finely tuned parameters. }
\end{center}
\end{figure}  

Since TCAF, without any magnetic fields, can explain bulk of the properties quite satisfactorily, we do not believe that magnetic fields 
play any major role other than perhaps collimating the jets and episodic supply of energies to outflows through re-connections
near disk surface. The advective flows can have dynamically unimportant fields and could also be responsible for a little viscosity that 
it may have. We believe that the field lines largely become toroidal and escape from the disk buoyantly \cite{cs94,sc94}. 
Several papers are being published in the literature with cartoons similar to CT95 configuration. In the so-called
magnetically arrested disks \cite{t11} the field lines are put by hand to do what we achieve using the natural centrifugal force.
In typical models of disks with a Compton cloud as in our CENBOL, the CENBOL just appears as a blotted gas 
(see, Fig. 6). Such a configuration is inconsistent and as Chakrabarti\cite{c97}
pointed out, unless the shock is located very far away, one single component disk cannot produce a hard state as the
soft photons from the disk are always sufficient to cool down the Compton cloud formed by itself. Similarly, there are many GRMHD code results
in the literature. To our knowledge, all these start with very finely tuned magnetic field configurations (such as split monopole or perfectly aligned field put by hand)
and the code often runs less than a complete dynamical time scale at the outer edge. Such results are outright misleading and can not possibly conclude anything about the 
black hole astrophysics.

A major outcome of the TCAF solution which has yet to be explored is the possibility of significant amount 
of nucleosynthesis in the CENBOL. Nucleosynthesis inside thick accretion disks 
(which are nothing but CENBOL in the language of TCAF paradigm) 
has been studied quite extensively in the literature \cite{cetal87,c88,j89,ah95,mc00}.

The same CENBOL temperature used to produce Comptonized 
spectra is indeed enough to have significant nuclear burning in the CENBOL, though mostly in hard states when the 
advective component is strong and the Keplerian disk component is weak. Some results in the context of over-production of several elements, including iron, nickel etc.
are already in the literature. The energetics of nucleosynthesis may cause un-predictable behaviour in the flow. This important issue requires further study.

The one and only assumption that the TCAF paradigm makes so far is that the injected matter at the outer edge is {\it not Keplerian} to begin with. 
In active galaxies or high mass X-ray binaries none would question this assumption. Questions are raised in low-mass X-ray binaries 
where Roche-Lobe overflow is {\it assumed} to be taking place.
However, since our single assumption explains more observations that any Keplerian disk assumption, we believe that the 
a part of the flow from the companion surface has to lose most of its angular momentum by stellar field lines 
before it is accreted. This intriguing prospect has to be looked into to understand the success of TCAF.

\end{document}